\documentclass[prb,twocolumn,showpacs,amsmath]{revtex4}
\usepackage{graphicx}
\usepackage{color}
\usepackage{tabularx}

\begin{document}
\title{Investigations on the relationship between $T_c$ and the superconducting gap under magnetic and non-magnetic impurity substitutions in YBa$_2$Cu$_3$O$_{7-\delta}$}

\author{M. Le Tacon$^{1,2}$, A. Sacuto$^{1,2}$, Y. Gallais$^{1,2}$, D. Colson$^{3}$, and A. Forget$^{3}$}
\address{$^{1}$Mat\'eriaux et ph\'enom$\grave{e}$nes Quantiques (CNRS UMR 7162), Universit\'e Paris 7, , 10 rue Alice Domon et Léonie Duquet, 75205 Paris Cedex 13, France\\
$^{2}$Laboratoire de Physique du Solide ESPCI, 10 rue Vauquelin, 75231 Paris, France\\
$^{3}$Service de Physique de l'Etat Condens\'{e}e, CEA-Saclay, 91191 Gif-sur-Yvette, France} 

\begin{abstract}
We report electronic Raman scattering measurements on optimally doped YBa$_{2}$Cu$_{3}$O$_{7-\delta}$ where Zn or Ni impurities have been substituted to Cu. Using Raman selection rules, we have probed the superconducting gap in the nodal and anti-nodal regions. We show that under impurity substitutions, the energy of the anti-nodal peak detected in the superconducting state is not related to the critical temperature $T_c$ and that signatures of superconductivity disappear in the nodal regions. Our experimental findings advocate in favor of gapless arcs around the nodes. The breakdown of the relationship between the anti-nodal gap amplitude and $T_c$ is discussed in terms of local superconducting gap and pseudogap.
\end{abstract}

\maketitle
\date\today

\section*{INTRODUCTION}
It is now well established that in the superconducting state of cuprates, the gap in the 
excitation spectra has $d_{x2-y2}$ symmetry~\cite{Tsuei_RMP00}. This gap is usually associated with the superconducting order parameter $\Delta_{sc}$, and, in the framework of BCS theory, its amplitude is expected to scale with $T_c$. An interesting way of exploring the link between the gap energy and $T_c$ in cuprates consists in substituting impurities on the copper sites of the CuO$_2$ planes~\cite{Tarascon_PRB88}, more specifically impurities such as Zn (non-magnetic) or Ni (magnetic), known for destroying $T_c$ without changing the doping level~\cite{Bobroff_PRL97, Bobroff_PRL99}. In this paper, we shall focus on the effects of such impurities on the electronic Raman response in the superconducting state of cuprates.
\par
Inelastic scattering of light by quasiparticles - Electronic Raman scattering (ERS) - has the unique ability of probing charge dynamics in different regions of the Fermi surface, namely around the principal $(\pi,0)(0,\pi)$ and diagonal $(\pi, \pi)$) directions of k-space in $B_{1g}$ geometry and B$_{2g}$ geometry respectively. In the superconducting state, the B$_{1g}$ geometry probes the anti-nodal regions where the amplitude of the {\sl d}-wave gap is maximum, and the B$_{2g}$ geometry probes the nodal regions, where the amplitude of the gap vanishes~\cite{Devereaux_PRL94}. Cubic and linear frequency dependence of the low energy B$_{1g}$ and B$_{2g}$ responses of a {\sl d}-wave superconductor are expected and were experimentally measured in many cuprates~\cite{Staufer_PRL92, Chen_PRL94, Kang_PRL96, Sacuto_PRB00, Gallais_PhysicaC04}.
\par
Previous ERS studies on magnetic impurities substitutions in cuprates (Ni in Y-123~\cite{Gallais_PRL02} or Fe in Bi-2212~\cite{Misochko_PRB99}) have shown that the B$_{1g}$ peak energy seems to be insensitive to impurity substitutions. The case of Zn substituted Y-123, with only one Zn concentration, led to contradictory results. In one study the B$_{1g}$ peak energy was found to follow $T_{c}$ \cite{Limonov_PRB01}, while in another study it was argued to collapse to zero \cite{Martinho_PRB04} when $T_{c}$ = 72 K.
Here, we report ERS measurements on a wide range of concentrations of magnetic Ni and non-magnetic Zn impurities. We compare their respective effects on the Raman responses in $B_{1g}$ and $B_{2g}$ geometries. We show that i) the energy of the $B_{1g}$ superconducting peak remains constant under magnetic and non-magnetic substitutions up to 3\%, in contrast to what is expected for the pair breaking peak in a conventional superconducting condensate, and ii) the superconducting $B_{2g}$ Raman response merges with the normal one as Zn and Ni are introduced in the CuO$_2$ planes.

\section{Experiment}

\begin{figure}[ptbh]
\begin{center}
\hspace{0mm}\includegraphics[width=0.80\columnwidth]{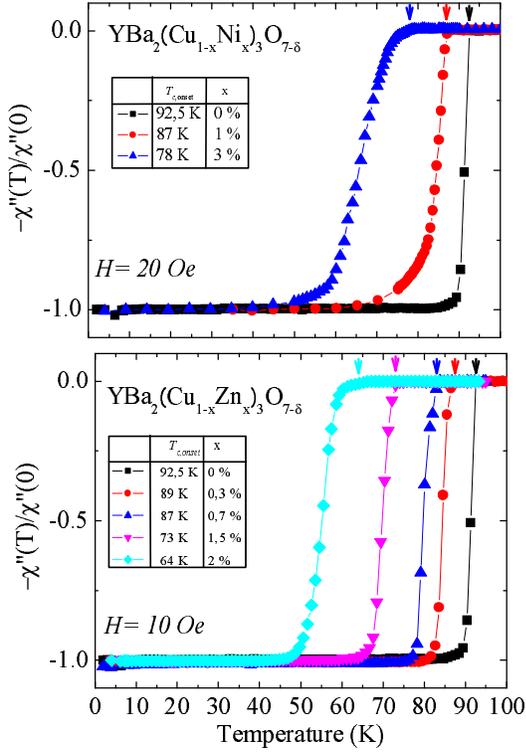}
\end{center}
\vspace{0mm}
\caption{(color online) dc magnetization in ZFC for various impurity concentrations YBa$_{2}$(Cu$_{1-x}$M$_{x}$)$_{3}$O$_{7-\delta }$ (M = Ni (upper panel) or Zn (lower panel)), The Tc on set are listed in the table for each concentration of impurities}.
\vspace{0mm}
\label{fig:TC}
\end{figure}

The synthesis protocol for YBa$_{2}$(Cu$_{1-x}$M$_{x}$)$_{3}$O$_{7-\delta }$ (M = Ni or Zn) single crystals have been previously detailed in ref.~\onlinecite{Gallais_PRL02} and~\onlinecite{LeTacon_PRB05}. The corresponding dc magnetization measurements are presented in fig.~\ref{fig:TC}.

We have labelled the pristine crystal Y-123, the Ni substituted crystals of ref.~\onlinecite{Gallais_PRL02} Y-123:Ni87K ($x_{Ni}\sim 1 \%$) and Y-123:Ni78K ($x_{Ni}\sim 3 \%$), and finally the Zn-substituted crystals of ref.~\onlinecite{LeTacon_PRB05} Y-123:Zn87K ($x_{Zn}\sim 0.3 \%$) , Y-123:Zn83K ($x_{Zn}\sim 0.7 \%$) , Y-123:Zn73K ($x_{Zn}\sim 1.5 \%$) and Y-123:Zn64K ($x_{Zn}\sim 2 \%$) respectively. The impurity concentrations have been checked by chemical analysis using a Castaing electron probe, and the $dT_c/x_{Zn}$ ($\sim$ 15K/\%) and $dT_c/x_{Ni}$ ($\sim$ 5K/\%) values are consistent with those previously reported~\cite{Bobroff_PRL99, Sidis_PRL00}.% Note that due to the Cu site in the CuO chains in YBCO, the effective in-plane impurity concentrations are given by 3/2*x.
\par
ERS have been carried out with a T64000 JY spectrometer in triple subtractive configuration. Crystals were mounted on the copper cold finger of an He circulation cryostat and cooled down to $10$~K, temperature at which all the superconducting spectra presented here have been measured (100 K for the normal state responses). The $514$~nm line of a $Ar^+, Kr^+$ laser was used. The laser power on the crystal surface was kept below $3$~mW to avoid any significant heating, which was smaller than $3$~K according to the Stokes-Anti-Stokes ratio. $B_{2g}$ and $B_{1g}$ geometries are obtained from cross polarizations of the incident and scattered electric fields along and at $45$ degrees of the Cu-O bounds of the CuO$_{2}$ layers. %Parallel polarizations at $45^{\mathrm{o}}$ from the Cu-O bounds have given the $A_{1g}$+$B_{2g}$ channel, and pure $A_{1g}$ channel was obtained by subtracting the $B_{2g}$ one from the $A_{1g}$+$B_{2g}$ one!
Raw spectra $I(\omega)$ have been corrected for the response of the spectrometer and the Bose factor $n(\omega ,T)$ to obtain the imaginary part of the Raman susceptibility $\chi^{\prime \prime }(\omega )\propto I(\omega )/[1+n(\omega ,T)]$.%Finally, in order to be able to compare the intensities of the different features on our spectra, they have all been normalized to unity at high energy ($\omega$ = 1000 $cm^{-1}$).

\section{Experimental results and data analysis}
\subsection{B$_{1g}$ response of Y-123}

\begin{figure}[ptbh]
\begin{center}
\hspace{0mm}\includegraphics[width=0.70\columnwidth]{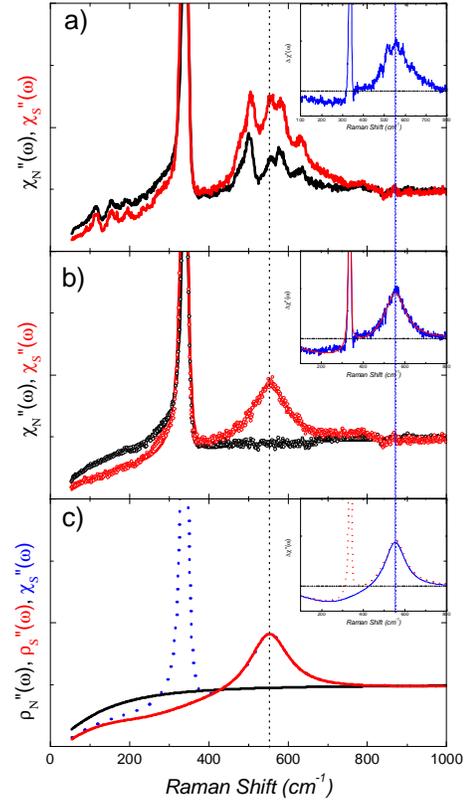}
\end{center}
\vspace{0mm}
\caption{(color online)a) B$_{1g}$ superconducting and normal Raman responses $\protect\chi^{\prime\prime}_{S}(\omega)$ (in red)and $\protect\chi^{\prime\prime}_{N}(\omega)$ (in black) of the pristine YBa$_{2}$Cu$_{3}$O$_{7-\protect\delta}$ crystal. The difference $\Delta\chi^{\prime\prime}(\omega) = \protect\chi^{\prime\prime}_{S}(\omega) - \protect\chi^{\prime\prime}_{N}(\omega)$ (in blue) is plotted in inset. b) Raman responses of a) after subtraction of weak phonon lines and responses calculated using the procedure described in the text. c) pure electronic contributions $\rho_N$ and $\rho_S$ to the Raman response extracted from the procedure described in the text.}
\vspace{0mm}
\label{fig:fit}
\end{figure}

In figure~\ref{fig:fit}-a) are shown the B$_{1g}$ Raman responses of the pristine Y-123 crystal in the normal (black curve) $\chi_{N}^{\prime \prime }(\omega )$ and superconducting (red curve) $\chi _{S}^{\prime \prime }(\omega )$ states. The normal state response consists of a flat electronic background with narrow phonon peaks superimposed. The strong peak at 340 $cm^{-1}$ corresponds to the out-of-phase motion of O(2,3) atoms along the c-axis. The 115 and 145 cm$^{-1}$ are respectively assigned to the motions of Cu and Ba along the c-axis and finally the 500 cm$^{-1}$ mode is related to the apical O(4) in-phase motion along c-axis. Additional weak features close to 200 cm$^{-1}$ and between 500 and 630 cm$^{-1}$ correspond to infrared forbidden modes due to slight CuO chains disorder~\cite{Macfarlane_PRB88, Liu_PRB88, Panfilov_PRB97}.
Entering into the superconducting state, we observe a redistribution of the spectral weight from the low energy region to the high one (400-800cm$^{-1}$). The weak phonon features are located inside the region where the background redistribution induced by superconductivity takes place. According to previous studies, their temperature dependence exhibit small energy shifts~\cite{Panfilov_PRB97} ($\sim 2$ cm$^{-1}$) as well as slight broadenings ($\lesssim 10\%$)\cite{Altendorf_PRB93}, that cannot explain alone the renormalization effect observed here).
Resonant Raman scattering measurements performed under ultra-violet excitation line \cite{Klein_JPCS06, Rubhausen_JPCS06} have recently unambiguously confirmed the electronic character of this redistribution.
\par
We have fitted the 100-200 cm$^{-1}$ and 500-630 cm$^{-1}$ groups of phonon lines using simple lorentzian profiles in the normal state and assuming a weak temperature dependence, as described above, in the superconducting state.  Figure~\ref{fig:fit}-b) exhibits the normal and superconducting Raman responses after subtraction of these phonon lines.
We note that due to the 340 cm$^{-1}$ phonon, the small orthorhombic distortion in Y-123~\cite{Nemetschek_EPB98} and the contribution from the Cu-O chains, the $w^3$ power law for the superconducting response of a {\it d}-wave superconductor~\cite{Devereaux_PRL94} is masked, contrary to other systems such as tetragonal Hg-based systems~\cite{Sacuto_PRB00, LeTacon_PRB05} where cubic law is clearly seen.
\par
In order to extract the 2$\Delta$ value from the Raman spectra free of weak phonon lines, we use a phenomenological fit of the electron-phonon coupled spectra in the normal and superconducting states (see refs.~\onlinecite{Klein_book}). The normal state susceptibility is given by :

\begin{equation}
\chi_{N}^{\prime \prime }(\omega ) = \rho_{N}(\omega) + \frac{\frac{S^2}{V^2}+2\rho_{N}(\omega)\epsilon S - V^2\rho_{N}(\omega)^2}{\Gamma(1+\epsilon^2)}
\label{eq:fit}
\end{equation}

where $\epsilon =\frac{\omega-\Omega}{\Gamma}$, $\Omega$ being the phonon energy ($\Omega$$\sim$ 340 cm$^{-1}$), $\Gamma$ its linewidth, and $S$ the ratio between phononic and electronic matrix elements. The three last parameters are renormalized by the electron-phonon coupling $V$ (a direct consequence of this coupling is the asymetric - Fano - lineshape of the B$_{1g}$ phonon line).

Following the same protocol as in ref.~\onlinecite{Limonov_PRB01}, we have accurately reproduced the electronic background in the normal state by using the expression 
$$\rho_N(\omega) = C\frac{\omega}{\sqrt{\omega+\omega_T}}$$ which is linear in frequency for $\omega \rightarrow 0$ and constant for $\omega \rightarrow \infty$ ($C$ and $\omega_T$ are fitting parameters).
To fit the Raman response $\chi_{S}^{\prime \prime }(\omega )$ in the superconducting state, we can also use eq.~\ref{eq:fit} replacing the electronic background $\rho_{N}(\omega)$ of the normal state by a renormalized $\rho_{S}(\omega)$ function. This was achieved by adding, as in ref.~\onlinecite{Limonov_PRB01}, two lorentzian profiles of opposite signs (one for the low energy loss of spectral weight, and one for the 400-800 cm$^{-1}$ enhancement) to $\rho_{N}(\omega)$. The result of this fitting procedure is displayed in Fig.~\ref{fig:fit}-b).
Fig.~\ref{fig:fit}-c), displays the pure electronic contributions $\rho_N(\omega)$ and $\rho_S(\omega)$ to the normal and superconducting responses deduced from these fits.
\par
The 2$\Delta$ value is assigned to the energy of the maximum of $\rho_S(\omega)$. It corresponds to 552 cm$^{-1}$ (8.6 $k_{B}T_{c}$) and is quite similar to the energy deduced from subtracting the normal state contribution to the superconducting of the raw spectra (2$\Delta$=556 cm$^{-1}$). These values are consistent with those previously reported~\cite{Chen_PRB93} in YBCO system. 
We have also added in dotted lines on Fig.~\ref{fig:fit}-c) the calculated electron-phonon coupled spectra in the superconducting state. We notice that the energy of the maximum is not altered by the electron-phonon coupling and can thus be directly extracted from the difference $\chi _{S}^{\prime \prime }(\omega )-\chi _{N}^{\prime \prime }(\omega )$ between the normal and superconducting responses.

\subsection{B$_{1g}$ response of Y-123 with impurities}

\begin{table}
\caption{\label{tab:fitparam} Fitting parameters}
\begin{ruledtabular}
\begin{tabular}{|c|c|c|c|c|}
 Sample & $\omega$ (cm$^{-1}$) & $\Gamma$ (cm$^{-1}$) & S & V\\
\hline
\hline
Y-123 & 338.6 & 6.82 & -6.99 & 0.79\\
\hline
\hline
Y-123:Ni87K & 337.8 & 9.37 & -3.86 & 0.543\\
\hline
Y-123:Ni78K & 339.2 & 8.25 & -4.25 & 0.542\\
\hline
\hline
Y-123:Zn87K & 337.4 & 9.27 & -2.81 & 0.346\\
\hline
Y-123:Zn83K & 337.1 & 9.32 & -2.39 & 0.330\\
\hline
Y-123:Zn73K & 338.9 & 9.27 & -3.28 & 0.447\\ 
\hline
Y-123:Zn64K & 338.8 & 7.92 & -4.29 & 0.549\\ 
\end{tabular}
\end{ruledtabular}
\end{table}

\begin{figure}[ptbh]
\begin{center}
\hspace{0mm}\includegraphics[width=0.85\columnwidth]{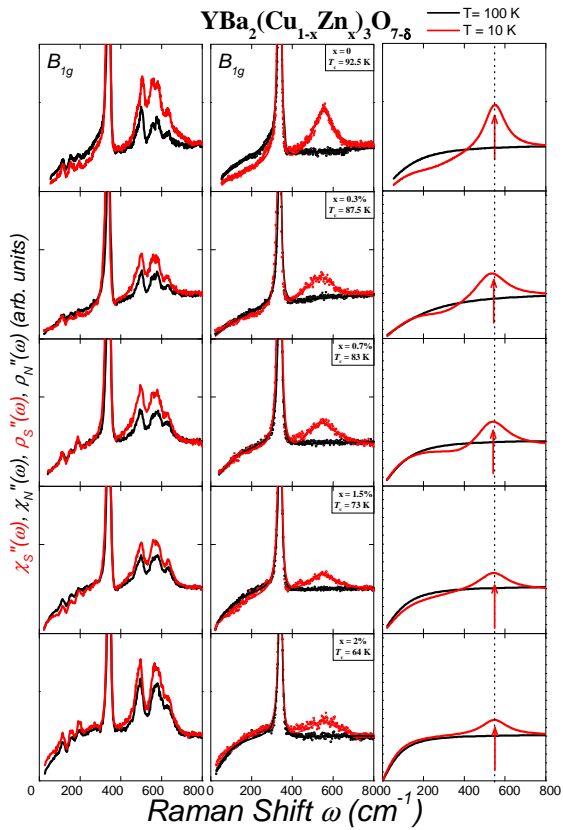}
\end{center}
\vspace{0mm}
\caption{(color online)(left panel) B$_{1g}$ superconducting and normal Raman responses $\protect\chi^{\prime\prime}_{S}(\omega)$ (in red) and $\protect\chi^{\prime\prime}_{N}(\omega)$ (in black) of the YBa$_{2}$(Cu$_{1-x}$Zn$_x$)$_{3}$O$_{7-\protect\delta}$ crystals. (Central panel) responses of the left panel after subtraction of weak phonons (dotted lines) and their fits (solid lines) obtained from the procedure described in the text. (Right panel) Pure electronic contributions: $\rho_{S}(\omega)$ and $\rho_{N}(\omega)$ in the superconducting (red) and normal state (black) respectively.}
\vspace{0mm}
\label{fig:Zn}
\end{figure}

\begin{figure}[ptbh]
\begin{center}
\hspace{0mm}\includegraphics[width=0.85\columnwidth]{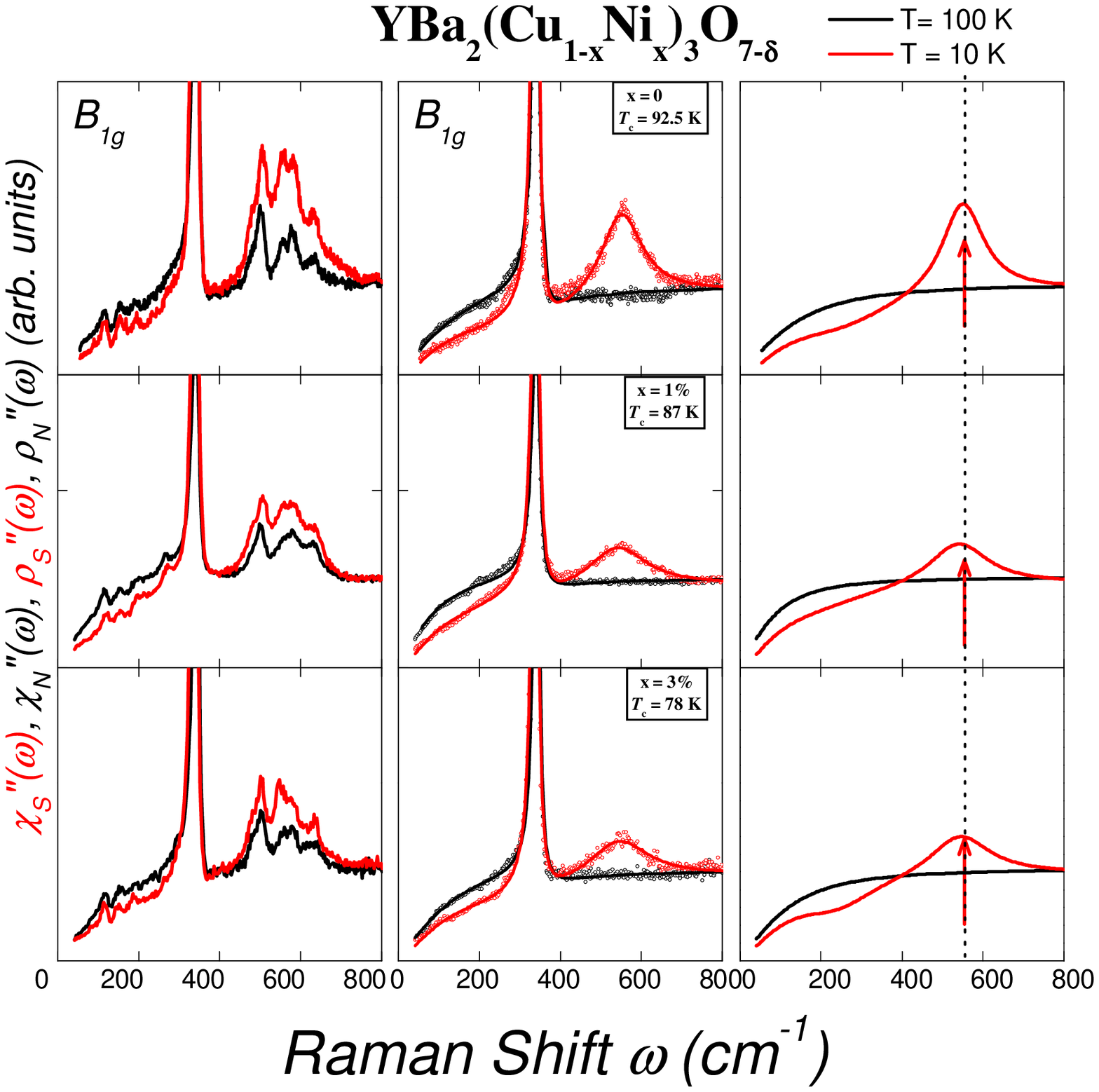}
\end{center}
\vspace{0mm}
\caption{(color online)(left panel) B$_{1g}$ superconducting and normal Raman responses $\protect\chi^{\prime\prime}_{S}(\omega)$ (in red) and $\protect\chi^{\prime\prime}_{N}(\omega)$ (in black) of the YBa$_{2}$(Cu$_{1-x}$Ni$_x$)$_{3}$O$_{7-\protect\delta}$ crystals. (central panel) responses of the left pannel after subtraction of weak phonons (dotted lines) and their fits (solid lines) obtained from the procedure described in the text. (Right panel) Pure electronic contributions: $\rho_{S}(\omega)$ and $\rho_{N}(\omega)$ in the superconducting (red) and normal state (black) respectively.}
\vspace{0mm}
\label{fig:Ni}
\end{figure}

Let us focus now on the B$_{1g}$ Raman spectra of Ni and Zn substituted Y-123 shown in Figures~\ref{fig:Ni} and~\ref{fig:Zn}, respectively. The left panels of the figures show the raw B$_{1g}$ spectra, the central ones show the same spectra after subtraction of the small phonon lines as well as their fits (solid lines) with the protocol described above, and finally, the right panels exhibit the normal $\rho_N(\omega)$ and superconducting $\rho_S(\omega)$ electronic contributions extracted from our analysis for each impurity concentration.
\par
The first striking feature is that Ni and Zn have qualitatively similar effects on the B$_{1g}$ Raman response of Y-123, with an expected~\cite{Devereaux_PRL95} broadening and a decrease of the superconducting B$_{1g}$ peak intensity. More surprisingly, we observe that the B$_{1g}$ peak energy neither collapses to zero as $T_{c}$ reaches 72 K (as claimed in ref.~\cite{Martinho_PRB04}), nor scales with $T_{c}$~\cite{Limonov_PRB01}, but \textbf{remains constant} over a wide concentration range and down to $T_{c}$ = 64 K.
\par
\begin{figure}[ptbh]
\begin{center}
\hspace{0mm}\includegraphics[width=0.82\columnwidth]{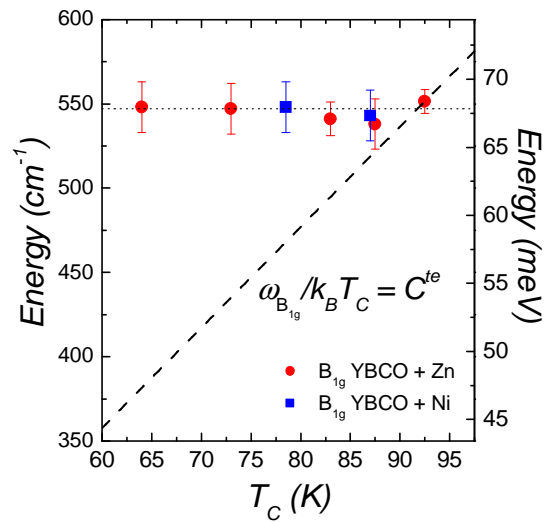}
\end{center}
\vspace{0mm}
\caption{%a) Spectral weight decrease of the B$_{1g}$ peak in the superconducting state under magnetic and non-magnetic impurity substitutions. To compare the Raman intensities between the different crystals, the high energy continuum have been normalized. The spectral weight corresponds to the integration of the $\Delta \chi^{\prime\prime}$ functions between 400 and 800 cm$^{-1}$. b) 
(color online) Energies of the B$_{1g}$ superconducting responses (in units of $T_c$) under magnetic Ni (squares) and non-magnetic Zn (circles) impurity substitutions.}
\vspace{0mm}
\label{fig:Bilan}
\end{figure}

 We have reported in figure~\ref{fig:Bilan} the different values of the pair breaking peak energies extracted from our data. The dashed line corresponds to the expected values of these energies assuming 2$\Delta/k_BT_c$ is kept constant. Clearly there is {\bf{no scaling between $\omega_{B_{1g}}$ and the critical temperature $T_c$}}.

\par

An independent confirmation of this fact can be inferred by looking more carefully at the fitting parameters used for the superconducting response (see table~\ref{tab:fitparam}).
We find a slight increase of the B$_{1g}$ phonon linewidth $\Gamma$ as well as a small decrease of both the electron-phonon coupling parameter $V$ and $\left|S\right|$ when impurities are added. These trends are opposite to those observed when the B$_{1g}$ peak energy decreases and gets closer to the 340 cm$^{-1}$ phonon as, for instance, in the case of overdoping~\cite{Chen_PRB93} or under pressure~\cite{Goncharov_JRS03}. In these cases, spectacular changes in the electron-phonon coupling strength lead to an important increase of the phonon linewidth which is not observed here.

\subsection{B$_{2g}$ response}

\begin{figure}[ptbh]
\begin{center}
%\hspace{0mm}
\includegraphics[width=0.75\columnwidth]{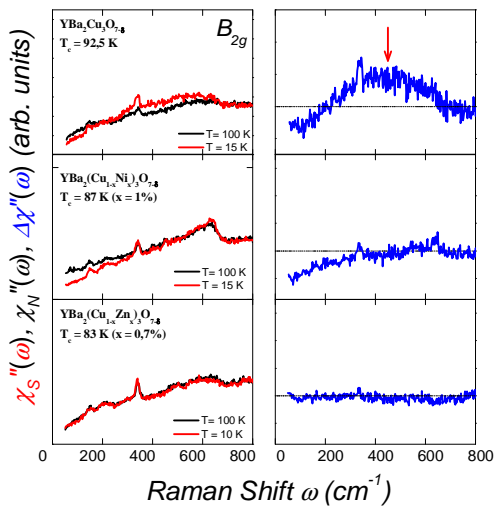}
\end{center}
%\vspace{0mm}
\caption{(color online) Superconducting and normal B$_{2g}$ Raman responses $\protect\chi^{\prime\prime}_{S}(\omega)$ (in red) and $\protect\chi^{\prime\prime}_{N}(\omega)$ (in black) of Y-123, Y-123:Ni87K and Y-123:Zn83K single crystals.
The differences $\Delta\chi^{\prime\prime}(\omega) = \protect\chi^{\prime\prime}_{S}(\omega) - \protect\chi^{\prime\prime}_{N}( \omega)$ (in blue) are plotted on the right panels.}
\vspace{0mm}
\label{fig:B2g}
\end{figure}

The nodal B$_{2g}$ responses for Y-123, Y-123:Ni87K and Y-123:Zn83K in the normal and the superconducting states are displayed in Fig.~\ref{fig:B2g}.  There are no Raman active phonons in this symmetry for tetragonal crystals, but due to the small orthorhombic distortion in the Y-123 family, some weak features are present in the spectra. %signatures of the 340 cm$^{-1}$  \textcolor{red}{je ne pense pas qu'ici le rappel de la symmetrie des phonons soit judicieux} and 120 cm$^{-1}$ phonons are present on the spectra. 
These contributions are marginal and the main part of the spectra is of electronic origin. In the pristine case, we observe renormalization of the B$_{2g}$ electronic continuum in the superconducting state. As expected theoretically for a {\sl d}-wave superconductor, we observe i) that the renormalization is much weaker than in the B$_{1g}$ case and ii) a linear dependence of the B$_{2g}$ response with frequency at low energy. In order to extract a characteristic energy from our data, we have subtracted the normal state contribution from the superconducting one (see right pannel of Fig.~\ref{fig:B2g}). This subtraction reveals a broad peak with a maximum around 470 cm$^{-1}$.
It appears that the nodal superconducting Raman response merges quickly with the normal one as impurities are inserted. Indeed, we detect a small loss of spectral weight at low energy for Y-123:Ni87K ($x_{Ni}$ = 1\%), but no difference between the superconducting and the normal $B_{2g}$ responses is observed for Y-123:Zn83K ($x_{Zn}$ = 0.7\%). In contrast to the B$_{1g}$ Raman response, we cannot extract from our measurements a $B_{2g}$ peak energy as function of impurity concentrations.

\section{Discussion and comparison to other experiments}

\subsection{Nodal Response}
The disappearance of any signature of the superconducting condensate in the the $B_{2g}$ Raman response as impurities are inserted is consistent with previous experiments which have shown that the $B_{2g}$ response is generally strongly affected by any scattering source (e.g. structural disorder)~\cite{Gallais_PRB05, Nemetschek_PRL97}. However, the effect reported here appears stronger than the one theoretically expected~\cite{Devereaux_PRL95}. Our experimental results are in better agreement with the appearance of gapless arcs around the nodes in presence of impurities for a gap of $d_{x2-y2}$ symmetry~\cite{Haas_PRB97}. Indeed, in this scenario, we do not expect to observe any difference between the superconducting and the normal B$_{2g}$ responses as impurities are introduced.

\subsection{Anti-Nodal Response}

The robustness of the B$_{1g}$ peak energy when impurities are inserted in the CuO$_2$ plane raises the question whether it is a genuine result or the consequence of an accidental cancellation between two antagonist effects. Indeed a small underdoping is expected to increase the B$_{1g}$ peak energy (ref. \onlinecite{Chubukov_PRB06}, \onlinecite{Le Tacon_NaturePhysics06}, \onlinecite{Devereaux_RMP07} and ref. therein) while at the same time impurity substitutions would induce a decrease of the peak energy. The two combined effects might then result in an apparent albeit fortuitous robustness of the peak energy.
\par
This scenario is however highly improbable because (i) the crystals have been annealed under oxygen pressure to be optimally doped (this is confirmed by the phonon peak locations), and (ii) because the changes of $T_c$ due to impurity substitution are much more drastic in the underdoped side~\cite{Bobroff_PRL99}. This would imply a strong disagreement between the impurity concentrations measured by the Castaing electron probe and the loss of $T_c$ according to the $dT_c/x_{Zn}$ ($\sim$ 15K/\%) and $dT_c/x_{Ni}$ ($\sim$ 5K/\%) ratios. This is not the case here. Another possible source for increasing the energy of the B$_{1g}$ peak~\cite{Devereaux_RMP07} is the presence of an anisotropic s-wave component in the superconducting gap. In this case a small increase of the energy of the superconducting peak is theoretically predicted~\cite{Devereaux_PRL95}, but this effect should be rather small and is unlikely to cancel out the energy decrease of the $\Delta$ peak~\cite{sasym}.

In the absence of any other potential source for an increase of the B$_{1g}$ peak energy we are led to conclude that the B$_{1g}$ peak energy is genuinely robust under impurity substitutions.
\par
Previous angle resolved photoemission spectroscopy (ARPES) experiments on electron-irradiated Bi-2212 samples (non-magnetic impurity effects in cuprates can be achieved by electron irradiation~\cite{RA2003}) revealed a similar robustness of the gap energy in the anti-nodal regions. Consistently with our findings $T_c$ was reduced to 62 K and no changes in the gap energy were observed~\cite{Vobornik_PRL01}. More recently, Takahashi's group has reported measurements of the angular dependence of the gap energy under magnetic and non-magnetic impurity substitutions\cite{Terashima_JPCS06}. The gap energy in the antinodal regions was shown to remain constant and gapless arcs around the nodes were observed, in agreement with the present findings.
\par
All these experiments are in agreement and they confirm that the antinodal peak energy is robust under impurity substitutions. This cannot be easily understood in the framework of conventional Abrikosov-Gorkov' theory. In this approach the presence of potential scatterers leads to a reduction of $T_c$ for a d-wave superconductor and is accompanied with a decrease of the superconducting order parameter amplitude $\Delta$~\cite{Fehrenbacher_PRB94, Haas_PRB97}, leaving the ratio between the gap and T$_c$ constant. If we relate the energy of the antinodal peak to the pair breaking energy ($\sim$ $\Delta$) as it is usually admitted~\cite{Devereaux_PRL94, Chen_PRL94, Staufer_PRL92, Sacuto_PRB00, Kang_PRL96, Gallais_PhysicaC04}, the B$_{1g}$ peak energy has to soften under impurity substitutions.
\par
At least two scenarios deserve to be explored in order to understand the robustness of the B$_{1g}$ superconducting peak, {\sl i. e.} the breakdown of the relationship between $T_c$ and $\Delta$, under impurity substitutions in YBCO.

The first one is built on the consensus that the B$_{1g}$ peak is directly related to the order parameter ($\omega_{B_{1g}}$ $\sim$ $2\Delta$) and based on local probe measurements showing that the superfluid condensate is only altered near the impurity site~\cite{Pan_Nature00, Hudson_Nature01}, on a length scale on the order of the coherence length $\xi$ (typically 15-20 \AA~ in optimally doped cuprates). This is the so-called "swiss cheese" picture, first inferred from superfluid density measurements~\cite{Nachumi_PRL96}. As pointed out by Franz {\sl et al.}~\cite{Franz_PRB97}, in this picture, the superconducting gap should be considered as a local quantity $\Delta(\vec r)$, affected only locally by impurity scattering, and cannot be replaced in the gap equation by its spatial average, which is a crucial condition for the validity of Abrikosov-Gorkov' approach.
It has been shown in this context that the decrease of $T_c$ was faster than the decrease of the average value of the superconducting gap. $\Delta(\vec r)$ vanishes on a length scale of $\xi$ around the impurity site, but remains unmodified with respect to the pristine case far from the impurity. To our knowledge, the theoretical Raman response of a superconductor with such inhomogeneities in the superconducting gap has not been calculated yet. However since the superconducting contribution to the response comes only from the regions far from impurity sites, we do not expect any shift of the pair breaking peak energy under impurity substitutions.

%The first one is based on the "swiss cheese model" where superconductivity is locally suppressed around each impurity site. In this scheme, the superfluid density at low temperature is proportional to Tc (confere to the Uemura plot, Mathieu mettre les references de Uemura). As impurities are introduced the superfluid density is reduce and thus Tc. However, the B$_{1g}$ superconducting peak related to superconducting domains out of the impurity site is not altered and therefore keeps constant its energy. \textcolor{red}{(Mathieu cette proposition m'a ete faite par Andy Millis qu'il faudra remercier dans les acknowledgements et est aussi en accord avec les manips de RMN de Julien)}.
\par
The second scenario raises the question of the nature of the B$_{1g}$ peak. It is indeed usually considered as the superconducting order parameter, but it is well-known that in several circumstances its energy does not scale with $T_c$. This is the case in the underdoped regime where the B$_{1g}$ peak energy increases continuously as $T_c$ decreases (refs. \onlinecite{Chubukov_PRB06}, \onlinecite{Le Tacon_NaturePhysics06}, \onlinecite{Devereaux_RMP07} and ref. therein) and under impurity substitutions, as shown by the present measurements.
In fact, the B$_{1g}$ peak energy behaves strikingly like the onset temperature $T^*$ of the pseudo-gap. $T^*$ increases as the doping decreases (see for instance ref.~\onlinecite{Timusk_RPP99} and ref. therein) and is also known to be insensitive to impurity substitutions~\cite{Alloul_PRL91, Yamamoto_PRB02}. Since near optimal doping $T_c$ and $T^*$ are very close, we are tempted to speculate that $T^*$ and $\omega_{B_{1g}}$ are related to the same energy scale, namely the pseudogap.
\par
Which of the two scenarios is at play in the cuprates is still an open question and cross-analysis between ARPES, Raman, tunneling and optical conductivity as well as further theoretical investigations are required to solve this important problem.

\section*{CONCLUSION}

We have presented the superconducting Raman responses in impurity substituted optimally doped YBa$_2$Cu$_3$O$_{7-\delta}$ single crystals. For both magnetic Ni and non-magnetic Zn substitutions, we found that the pair breaking peak energy measured in the antinodal regions remains constant, in contrast with Abrikosov-Gorkov' calculations for a conventional BCS superfluid condensate. In order to explain our findings, we have explored two scenarios: the first one takes into account the local inhomogeneity of the superconducting gap and the second one links the antinodal peak energy to the pseudo-gap energy rather than the superconducting order parameter. In the nodal regions, no signature of superconductivity is observed in presence of impurities, suggesting the apparition of gapless arcs around the nodes.

\acknowledgments
We are gratefull to E. Sherman, J. Bobroff, H. Alloul, S. Paihl\`es, Y. Sidis, Ph. Bourges, M. Cazayous, K. Behnia, G. Deutscher, Ph. Monod and A.Millis for very fruitful discussions.


\begin{thebibliography}{0}
\bibitem{sasym} The calculation of the Raman response for an anisotropic s-wave superconductor presented in ref.~\onlinecite{Devereaux_PRL95} shows an increase of about 10\% of the energy of the pair breaking peak in B$_{1g}$ symmetry with $\Gamma/\Delta = n_{Imp}/(\pi N_F)$ = 0.2. With $N_F \sim$ 1eV$^{-1}$, it corresponds to an impurity level of 2.5\% per CuO$_2$ plane, that would lead to a decrease of $T_c$ of about 25 K, {\sl i.e.} almost 30\% of the $T_c$ for the pristine sample. In the case of YBCO, only a small s-wave component in the superconducting gap exists, the energy shift expected is thus marginal.
\end{thebibliography}
\end{document}